\documentclass{article}
\usepackage{spconf,amsmath,graphicx,hyperref}

\usepackage{cite}
\usepackage{overpic}
\usepackage{amsmath,amssymb,amsfonts}
\usepackage{graphicx}
\usepackage{textcomp}
\usepackage{xcolor}
\usepackage{float}
\usepackage{amsthm}
\usepackage{graphicx}
\usepackage{epstopdf}
\usepackage{amsmath,bm}
\usepackage{amsfonts}
\usepackage{amssymb}
\usepackage{color}
\usepackage{subfigure}
\usepackage{multirow}
\usepackage{multicol}
\usepackage{url}
\usepackage{soul,xcolor}
\usepackage{algorithm}
\usepackage{algpseudocode}

\theoremstyle{plain}

\newtheorem{remark}{Remark}

\newcommand{\vect}[1]{\mathbf{#1}}

\def\Htran{\mbox{\tiny $\mathrm{H}$}}
\def\Ttran{\mbox{\tiny $\mathrm{T}$}}
\def\CN{\mathcal{N}_{\mathbb{C}}}

\title{Network-controlled repeaters under power amplifier non-linearities}

\name{Özlem Tuğfe Demir$^*$ and Emil Björnson$^\dagger$\thanks{This work was supported by the Knut and Alice Wallenberg Foundation.}}
\address{$^*$Department of Electrical and Electronics Engineering, Bilkent University, Ankara, Turkiye  \\
$^\dagger$Department of Communication Systems, KTH Royal Institute of Technology, Stockholm, Sweden \\
Email: ozlemtugfedemir@bilkent.edu.tr, emilbjo@kth.se
}

\begin{document}
\ninept

\maketitle

\begin{abstract}
Network-controlled repeaters (NCRs) are a low-cost means to extend coverage and strengthen macro diversity in wireless networks. They operate in real time by amplifying and re-transmitting the incoming signal with only hardware-level delays, without requiring any channel state information (CSI) at the repeater itself. However, their power amplifiers (PAs) generate non-linear distortion that is jointly forwarded with the desired signal and can undermine multiuser performance unless the distortion statistics are exploited. This paper develops a distortion-aware (DA) uplink framework for repeater-assisted massive MIMO (RA-MIMO) under PA non-linearities. We adopt a memoryless third-order polynomial model for the repeater PA and characterize the achievable spectral efficiency (SE) using the Bussgang decomposition. Closed-form expressions are derived for the Bussgang gain matrix and the distortion covariance. We also design a DA combining vector that maximizes the effective signal-to-interference-plus-distortion ratio. 
\end{abstract}

\begin{keywords}Network-controlled repeater, hardware impairments, power amplifier non-linearities.
\end{keywords}

\section{Introduction}
\label{sec:intro}

Massive multiple-input multiple-output (MIMO) provides substantial beamforming gains thanks to a high number of antennas and spatial multiplexing by serving many user equipments (UEs) on the same time–frequency resources. However, in conventional cellular deployments, cell-edge UEs often experience weak channel gains due to pathloss and blockage \cite{bai2025repeater}. A commonly proposed remedy is cell-free massive MIMO, which densifies the network with many low-cost access points (APs), but large-scale synchronization, fronthaul/backhaul, and site-deployment challenges remain major bottlenecks.

An alternative low-cost, real-time relaying approach is the \emph{network-controlled repeater} (NCR), standardized in 3GPP Release~18 \cite{wen2024shaping}. NCRs can be leveraged as active scatterers to enhance macro diversity within a cell, giving rise to repeater-assisted massive MIMO (RA-MIMO) \cite{willhammar2025achieving,topal2025fair}. Unlike distributed APs, repeaters do not require wired backhaul and impose far less stringent phase-synchronization requirements \cite{willhammar2025achieving}.

An NCR amplifies and re-transmits the received signal almost instantaneously—with a processing delay on the order of a few hundred nanoseconds \cite{willhammar2025achieving}—akin to full-duplex relays. Different from conventional amplify-and-forward relaying \cite{sanguinetti2012tutorial}, the repeater-aided and direct paths are superposed over the air in a single received signal expression. It is a practical alternative to reconfigurable intelligent surface (RIS)-based solutions, which typically require very large apertures to overcome the cascaded path loss.

\subsection{Related Works}

Repeaters are a long-established concept: commercial deployments began in the 2G era to improve coverage, with early studies documenting several deployment scenarios \cite{patwary2005capacity,tsai2010capacity,bai2025repeater}. In 5G New Radio (NR), substantial effort within 3GPP has focused on standardizing NCRs \cite{3GPPNCR}, and a growing body of work now examines their capabilities and use cases \cite{carvalho2025network,ayoubi2022network}.

Despite this progress, the impact of hardware impairments on repeater-assisted transmission remains insufficiently understood. For example, in vehicular settings, repeaters improve in-car coverage by relaying signals via a roof pick-up antenna but also forward amplified uplink noise and introduce additional interference such as passive intermodulation in the indoor antenna and leakage due to limited isolation between indoor and outdoor antennas \cite{10848174}.

\subsection{Contributions}

In this paper, we focus on distortion originating from power-amplifier (PA) non-linearities, unlike \cite{10848174} that analyzes vehicular repeaters and uplink interference mechanisms such as amplified noise, direct window paths, passive intermodulation, and insufficient antenna isolation. These non-linearities render the repeater output distorted and must be modeled explicitly; consequently, distortion-aware (DA) receiver processing is crucial \cite{bjornson2018hardware,9072380}. A key implication is that the achievable spectral efficiency (SE) depends non-linearly on the amplification factor: while moderate gains improve SE, excessive amplification forwards distortion more strongly, leading to an optimal operating point. We consider a single-cell uplink where a multi-antenna base station (BS) serves multiple UEs on the same time–frequency resources with assistance from a repeater. While the repeater amplifies a noisy superposition of the UE signals, it also introduces PA distortion, which we model via a memoryless polynomial non-linearity. We develop a Bussgang-based analytical framework for uplink RA-MIMO under PA nonlinearities, explicitly accounting for noise amplification at the repeater, thereby distinguishing our model from prior single-hop non-linear low-noise amplifier analyses \cite{bjornson2018hardware}. We derive the Bussgang gain matrix and the distortion covariance. Then, we design a DA combining vector that maximizes the resulting signal-to-interference-plus-distortion ratio.

\section{System Model}

We consider the uplink transmission from $K$ single-antenna UEs to a multi-antenna BS with $M$ antennas through a single-antenna NCR. The received signal at the repeater is \cite{topal2025fair}
\begin{align}
    \tilde{u} = \sqrt{p}\sum_{i=1}^Kh_is_i+n,
\end{align}
where $h_i \in \mathbb{C}$ is the channel coefficient between UE $i$ and the NCR, $p>0$ is the uplink transmit power of each UE, and $s_i\sim \CN(0,1)$ is the zero-mean unit variance data symbol of UE $i$, and  $n\sim \CN(0,\sigma_{\rm r}^2)$ is the independent receiver noise.  The repeater will amplify and re-transmit the received signal $\tilde{u}$. The linearly amplified signal without any PA distortion is 
\begin{align}
    u = \alpha \tilde{u} = \sqrt{p}\alpha\sum_{i=1}^Kh_is_i+ \alpha n,
\end{align}
where $\alpha>0$ is the amplification factor.

In this paper, we consider hardware impairments at the NCR side, and model the power amplifier non-linearities with the standard third-order polynomial distortion model \cite{9072380}. In addition to the ideally amplified signal $u=\alpha\tilde{u}$, there will be an additional third-order term multiplied by the compression parameter $\rho\leq 0$. The distorted output signal of the PA is
\begin{align}
    r = u+ \rho  |u|^2u.
\end{align}
Denoting the channel from the NCR to the BS by $\vect{g}\in \mathbb{C}^M$ and the direct channel from the UE $i$ to the BS by $\overline{\vect{h}}_i\in \mathbb{C}^M$, the received signal $\vect{y} \in \mathbb{C}^M$ at the BS is written as
\begin{align}
    \vect{y} = \vect{g}r+\sqrt{p}\sum_{i=1}^K\overline{\vect{h}}_is_i+\vect{w} ,
\end{align}
where $\vect{w}\sim \CN(\vect{0}, \sigma_{\rm BS}^2\vect{I}_M)$ is the independent receiver noise. 

\vspace{-2mm}

\section{Computation of Bussgang gain and distortion covariance matrix}

The Bussgang decomposition provides a principled approach for modeling memoryless non-linearities \cite{demir2020bussgang}. Specifically, it yields an exact probabilistic characterization in which the output of a non-linear function is represented as a linear scaling of the input combined with an additive distortion component that is uncorrelated with the input.

In this section, we compute the Bussgang decomposition with the goal of obtaining the achievable SEs, which are then provided in Section~\ref{sec:SE}.
Using the Bussgang decomposition
 \cite{demir2020bussgang}, the received signal at the BS can be decomposed as
 \begin{align}
 \vect{y} = \vect{B}\vect{s}+\boldsymbol{\eta}   , 
 \end{align}
 where $\vect{B}\in \mathbb{C}^{M\times K}$ is the Bussgang gain, $\vect{s} = [s_1 \, \ldots \, s_K]^{\Ttran}$ is the UEs' symbol vector, and $\boldsymbol{\eta}\in \mathbb{C}^{M}$ is the distortion, which is uncorrelated with $\vect{s}$ by construction. The Bussgang gain matrix is computed as \cite[Eqn. (20)]{demir2020bussgang}
 \begin{align}
 \vect{B} = \vect{C}_{ys}\vect{C}_s^{-1}
 \end{align}
 where $\vect{C}_{ys} = \mathbb{E}\{\vect{y}\vect{s}^{\Htran}\}$ is the cross-correlation matrix and $\vect{C}_s=\mathbb{E}\{\vect{s}\vect{s}^{\Htran}\}$ is the auto-correlation matrix of the symbol vector $\vect{s}$. Using the independence of the unit-variance data symbols, we have $\vect{C}_s=\vect{I}_K$ and this leads to 
 \begin{align}
     \vect{B} = \vect{C}_{ys}= \mathbb{E}\{\vect{y}\vect{s}^{\Htran}\}.
 \end{align}

 To compute the Bussgang gain matrix, we focus on the $(m,k)$th entry of it, and compute 
\begin{align}
    \left[\vect{B}\right]_{m,k} = \mathbb{E}\{y_ms_k^*\}
\end{align}
where  $y_m$ is the $m$th entry of $\vect{y}$. Denoting the $m$th entry of $\vect{g}$, $\overline{\vect{h}}_i$, and $\vect{w}$ by $g_m$, $\overline{h}_{i,m}$, and $w_m$, we obtain $y_m$ as
\begin{align}
   y_m &= g_mr + \sqrt{p}\sum_{i=1}^K\overline{h}_{i,m}s_i+w_m \nonumber\\
   & = g_mu +\rho g_m|u|^2u+ \sqrt{p}\sum_{i=1}^K\overline{h}_{i,m}s_i+w_m  \nonumber\\
   & =  g_m\alpha\tilde{u}+\rho g_m\alpha^3|\tilde{u}|^2\tilde{u}+ \sqrt{p}\sum_{i=1}^K\overline{h}_{i,m}s_i+w_m . \label{eq:ym}
\end{align}
Then, the $(m,k)$th entry of the $\vect{B}$ can be computed as 
\begin{align}
    &\left[\vect{B}\right]_{m,k} = \mathbb{E}\{y_ms_k^*\} \nonumber\\
    & =\mathbb{E}\Bigg\{\Bigg(g_m\alpha\tilde{u}+\rho g_m\alpha^3|\tilde{u}|^2\tilde{u}+ \sqrt{p}\sum_{i=1}^K\overline{h}_{i,m}s_i+w_m\Bigg)s_k^*\Bigg\} \nonumber\\
    & = g_m\alpha\mathbb{E}\big\{\tilde{u}s_k^*\big\} +\rho g_m\alpha^3\mathbb{E}\big\{|\tilde{u}|^2\tilde{u}s_k^*\big\}+\sqrt{p}\overline{h}_{k,m}. \label{eq:Bmk}
\end{align}
The remaining expectations in \eqref{eq:Bmk} can be computed as
\begin{align}
&\mathbb{E}\big\{\tilde{u}s_k^*\big\} = \mathbb{E}\Bigg\{\Bigg(\sqrt{p}\sum_{i=1}^Kh_is_i+n\Bigg)s_k^*\Bigg\}=\sqrt{p}h_k, \label{eq:exp1} \\
&\mathbb{E}\big\{|\tilde{u}|^2\tilde{u}s_k^*\big\}=\mathbb{E}\Bigg\{\Bigg|\sqrt{p}\sum_{i=1}^Kh_is_i+n\Bigg|^2\Bigg(\sqrt{p}\sum_{j=1}^Kh_js_j+n\Bigg)s_k^*\Bigg\} \nonumber \\
&=2p\sqrt{p}|h_k|^2h_k+2p\sqrt{p}h_k\sum_{i=1,i\neq k}^K|h_i|^2+2\sqrt{p}\sigma_{\rm r}^2h_k. \label{eq:exp2}
\end{align}
Inserting \eqref{eq:exp1}-\eqref{eq:exp2} into \eqref{eq:Bmk}, we obtain
    \begin{align}
    &\left[\vect{B}\right]_{m,k} =g_m\alpha\sqrt{p}h_k + 2\rho g_m \alpha^3 p\sqrt{p}|h_k|^2h_k \nonumber\\
    &+2\rho g_m \alpha^3 p\sqrt{p}h_k\sum_{i=1,i\neq k}^K|h_i|^2+2\rho g_m \alpha^3\sqrt{p}\sigma_{\rm r}^2h_k+\sqrt{p}\overline{h}_{k,m}.
\end{align}

After computing the Bussgang gain matrix $\vect{B}$, we can also compute the auto-correlation matrix of the distortion $\boldsymbol{\eta}$ by the uncorrelatedness of $\vect{s}$ and $\boldsymbol{\eta}$ as \cite[Eqn. (24)]{demir2020bussgang}
\begin{align}
 \vect{C}_{\eta} = \vect{C}_y-\vect{B}\vect{B}^{\Htran}
\end{align}
where $\vect{C}_{\eta}$ and $\vect{C}_y$ are the auto-correlation matrices of $\boldsymbol{\eta}$ and $\vect{y}$, respectively. Using \eqref{eq:ym}, the $(m,n)$th entry of $\vect{C}_y$ can be computed as
\begin{align}
&\left[\vect{C}_{y}\right]_{m,n} = \mathbb{E}\{y_my_n^*\} \nonumber \\ &=\mathbb{E}\Bigg\{\Bigg(g_m\alpha\tilde{u}+\rho g_m\alpha^3|\tilde{u}|^2\tilde{u}+ \sqrt{p}\sum_{i=1}^K\overline{h}_{i,m}s_i+w_m\Bigg ) \nonumber\\
&\quad \times \Bigg(g_n^*\alpha\tilde{u}^*+\rho g_n^*\alpha^3|\tilde{u}|^2\tilde{u}^*+ \sqrt{p}\sum_{j=1}^K\overline{h}_{j,n}^*s_j^*+w_n^*\Bigg)\Bigg\} \nonumber\\
& = g_mg_n^*\alpha^2\mathbb{E}\left\{|\tilde{u}|^2\right\}+\rho^2g_mg_n^*\alpha^6\mathbb{E}\left\{|\tilde{u}|^6\right\}+2\rho g_m g_n^*\alpha^4\mathbb{E}\left\{|\tilde{u}|^4\right\} \nonumber \\
&\quad +p\sum_{i=1}^Kh_{i,m}h_{i,n}^*+\sigma^2_{\rm BS}\delta[m-n] \nonumber\\
&\quad + g_m \alpha \mathbb{E}\Bigg\{\Bigg( \tilde{u}\sqrt{p}\sum_{j=1}^K\overline{h}_{j,n}^*s_j^*\Bigg)\Bigg\} \nonumber\\
&\quad + g_n^* \alpha \mathbb{E}\Bigg\{\Bigg( \tilde{u}^*\sqrt{p}\sum_{i=1}^K\overline{h}_{i,m}s_i\Bigg)\Bigg\} \nonumber\\
& \quad + \rho g_m \alpha^3 \mathbb{E}\Bigg\{\Bigg( |\tilde{u}|^2\tilde{u}\sqrt{p}\sum_{j=1}^K\overline{h}_{j,n}^*s_j^*\Bigg)\Bigg\} \nonumber\\
&\quad + \rho g_n^* \alpha^3 \mathbb{E}\Bigg\{\Bigg( |\tilde{u}|^2\tilde{u}^*\sqrt{p}\sum_{i=1}^K\overline{h}_{i,m}s_i\Bigg)\Bigg\}, \label{eq:Cym}
\end{align}
where $\delta[m-n]=1$ if $m=n$ and zero otherwise. The expectations given above are computed as
\begin{align}
&\mathbb{E}\left\{|\tilde{u}|^2\right\} = p\sum_{i=1}^K|h_i|^2+\sigma_{\rm r}^2,  \label{eq:exp3} \\
&\mathbb{E}\left\{|\tilde{u}|^4\right\} = 2\left(p\sum_{i=1}^K|h_i|^2+\sigma_{\rm r}^2\right)^2, \\
&\mathbb{E}\left\{|\tilde{u}|^6\right\} = 6\left(p\sum_{i=1}^K|h_i|^2+\sigma_{\rm r}^2\right)^3, \\
&\mathbb{E}\Bigg\{\Bigg( \tilde{u}\sqrt{p}\sum_{j=1}^K\overline{h}_{j,n}^*s_j^*\Bigg)\Bigg\}= p \sum_{i=1}^Kh_i\overline{h}_{i,n}^*, \\
&\mathbb{E}\Bigg\{\Bigg( \tilde{u}^*\sqrt{p}\sum_{i=1}^K\overline{h}_{i,m}s_i\Bigg)\Bigg\}=p \sum_{i=1}^Kh_i^*\overline{h}_{i,m},\\
& \mathbb{E}\Bigg\{\Bigg( |\tilde{u}|^2\tilde{u}\sqrt{p}\sum_{j=1}^K\overline{h}_{j,n}^*s_j^*\Bigg)\Bigg\}=2p^2\sum_{i=1}^K|h_i|^2h_i\overline{h}_{i,n}^* \nonumber\\
&\quad+2p^2\sum_{i=1}^K|h_i|^2\sum_{j=1,j\neq i}^Kh_j\overline{h}_{j,n}^*+2p\sigma_{\rm r}^2\sum_{i=1}^Kh_i\overline{h}_{i,n}^*, \\
& \mathbb{E}\Bigg\{\Bigg( |\tilde{u}|^2\tilde{u}^*\sqrt{p}\sum_{i=1}^K\overline{h}_{i,m}s_i\Bigg)\Bigg\}=2p^2\sum_{i=1}^K|h_i|^2h_i^*\overline{h}_{i,m} \nonumber\\
&\quad+2p^2\sum_{i=1}^K|h_i|^2\sum_{j=1,j\neq i}^Kh_j^*\overline{h}_{j,m}+2p\sigma_{\rm r}^2\sum_{i=1}^Kh_i^*\overline{h}_{i,m}.\label{eq:exp4}
\end{align}
Inserting \eqref{eq:exp3}-\eqref{eq:exp4} into \eqref{eq:Cym}, we obtain the entries of $\vect{C}_{y}$.

\subsection{Achievable Spectral Efficiency}\label{sec:SE}
Building on the Bussgang decomposition derived above, we now characterize the achievable SE. The BS applies receive combining vector $\vect{v}_k\in\mathbb{C}^M$ to the received signal $\vect{y}$ to decode the information signal of UE $k$ and obtains
\begin{align}
\vect{v}_k^{\Htran}\vect{y} = \vect{v}_k^{\Htran}\vect{b}_ks_k+\sum_{i=1,i\neq k}^K \vect{v}_k^{\Htran}\vect{b}_is_i + \vect{v}_k^{\Htran}\boldsymbol{\eta},
\end{align}
where $\vect{b}_k$ is the $k$th column of $\vect{B}$. Noting that the interference plus distortion is uncorrelated with the desired signal $\vect{v}_k^{\Htran}\vect{b}_ks_k$, we can use the worst-case uncorrelated additive noise theorem \cite{hassibi2003much} to compute a lower bound to the capacity for UE $k$, which is the achievable SE 
\begin{align}
\mathrm{SE}_k = \log_2\left(1+\frac{\left|\vect{v}_k^{\Htran}\vect{b}_k\right|^2}{\sum_{i=1,i\neq k}^K \left|\vect{v}_k^{\Htran}\vect{b}_i\right|^2+\vect{v}_k^{\Htran}\vect{C}_{\eta}\vect{v}_k }\right).
\end{align}
For each UE $k$, the optimal distortion-aware (DA) receive combiner $\vect{v}_k$ that maximizes the achievable SE, is given as
\begin{align}
\vect{v}_k^{\star} = \left( \sum_{i=1,i\neq k}^K\vect{b}_i\vect{b}_i^{\Htran}+\vect{C}_{\eta}\right)^{-1}\vect{b}_k \label{eq:DA}
\end{align}
leading to the SE
\begin{align}
\mathrm{SE}_k^{\star} =\log_2\left(1+ \vect{b}_k^{\Htran}\left( \sum_{i=1,i\neq k}^K\vect{b}_i\vect{b}_i^{\Htran}+\vect{C}_{\eta}\right)^{-1}\vect{b}_k\right). 
 \end{align}

\begin{remark}
While in this work, we use the closed-form Bussgang gain and distortion covariance primarily to characterize the achievable SE and design a DA combining vector, these analytical expressions are also valuable for future optimization tasks. In particular, they provide the tractable foundation needed to optimize system parameters such as the UEs’ transmit powers or the repeater amplification factor $\alpha$, which cannot be done reliably without explicit closed-form models of the distortion statistics.
\end{remark}

\vspace{-4mm}

\section{Numerical Results}

\vspace{-2mm}

In this section, we quantify the SE gains enabled by
(i) introducing a single-antenna NCR with amplification factor \(\alpha\) in the coverage area and
(ii) employing the DA receive combiner in \eqref{eq:DA},
relative to a distortion-unaware (DuA) combiner, which ignores the distortion statistics by forming \(\vect{v}_k\) under the (incorrect) assumption of an ideal PA (\(\rho=0\)). It is important to note that both DA and DuA schemes are evaluated under the same non-ideal PA model; the distinction is solely in the computation of the receive combining vectors.

We consider a BS equipped with \(M=64\) antennas serving \(K\) UEs over a bandwidth of \(B=100\)~MHz. Each UE transmits with power \(p=200\)\,mW. Thermal noise is included both at the repeater input and at the BS receiver using a noise power density of \(-204\)~dBW/Hz and a 5~dB noise figure, leading to $\sigma^2_{\rm BS}=\sigma^2_{\rm r}=-89$\,dBm.

\textbf{Geometry.} The BS is placed at the origin in the horizontal plane, i.e., \((x_{\mathrm{BS}},y_{\mathrm{BS}})=(0,0)\). A single-antenna repeater is located at \((x_{\mathrm{rep}},y_{\mathrm{rep}})=(200,0)\) meters. The UE positions are drawn independently and uniformly from the rectangle
\begin{align}
(x,y)\in [200,300]\times[-50,50]\ \text{meters},
\end{align}
so that each realization places all UEs in front of the BS and to the right of the repeater along the \(x\)-axis. The heights are fixed to \(z_{\mathrm{BS}}=25\)~m, \(z_{\mathrm{rep}}=15\)~m, and \(z_{\mathrm{UE}}=1.5\)~m, yielding explicit 3D coordinates \((x,y,z)\) for all nodes.

\textbf{Channel model.} Large-scale fading is modeled via a non-line-of-sight pathloss formula $-34.53-38\cdot\log_{10}(d)$ \cite{3GPP25996}, where $d$ is the distance in meters, while the small-scale fading is independent and identically distributed Rayleigh across links and realizations.

 To assess the repeater-assisted transmission, we sweep the repeater amplification \(\alpha\) and the non-linear compression parameter \(\rho\); here, \(\rho=0\) corresponds to an ideal PA, whereas \(\rho<0\) captures increasing compression non-linearity. For each network realization, we compute the sum SE under both DA and DuA combining; the DuA combiner assumes \(\rho=0\) when forming \(\vect{v}_k\), while the DA combiner accounts for the actual \(\rho\). Results are averaged over 100 independent channel setups.

\begin{figure}[t!]
        \centering
	\begin{overpic}[width=0.98\columnwidth,trim=0.2cm 0cm 0.5cm 0.5cm,clip,tics=10]{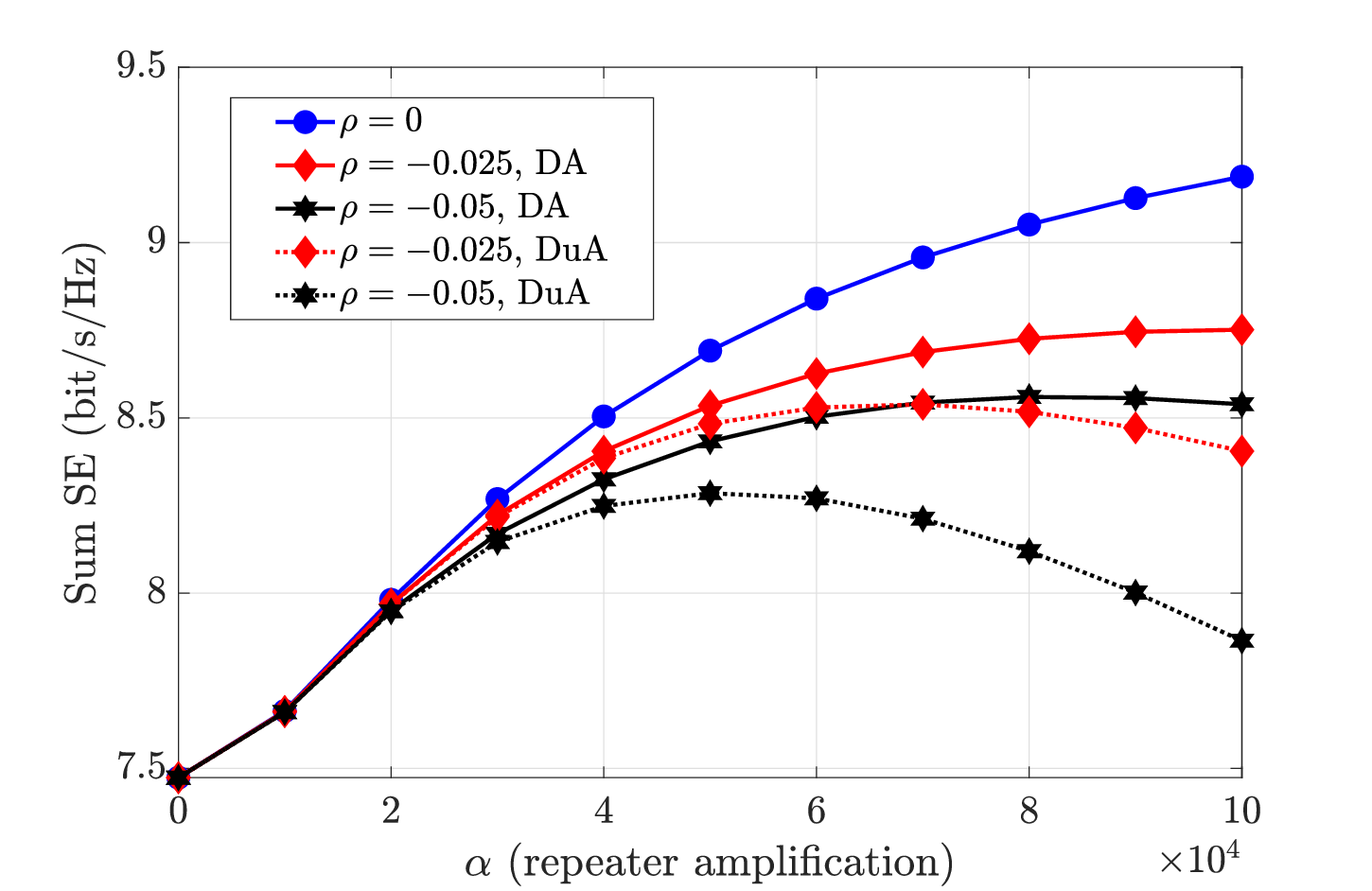}
\end{overpic} 
\vspace{-5mm}
        \caption{Sum SE in terms of repeater amplification gain $\alpha$ for $K=4$, when the repeater-BS distance is $200$\,m and UEs occupy a $100\,\text{m}\times100\,\text{m}$ region. }
        \label{fig1}
        \vspace{-4.2mm}
\end{figure}

In Fig.~\ref{fig1}, we consider \(K=4\) UEs and sweep the repeater amplification from \(\alpha=0\) to \(\alpha=10^{5}\). The blue curve corresponds to an ideal power amplifier, i.e., \(\rho=0\). As the non-linearity worsens (more negative \(\rho\)), the SE decreases; the loss is substantially larger with DuA combining, which ignores the distortion statistics captured via the Bussgang decomposition. In contrast, the proposed DA combining remains consistently more robust. Importantly, relative to the no-repeater baseline (\(\alpha=0\)), repeater-assisted transmission achieves higher SE over a broad range of \(\alpha\). However, without DA combining, excessive amplification can exacerbate non-linear distortion and ultimately degrade SE, which is caused by the joint amplification of noise and non-linear distortion.

\begin{figure}[t!]
        \centering
	\begin{overpic}[width=0.98\columnwidth,trim=0.2cm 0cm 0.5cm 0.5cm,clip,tics=10]{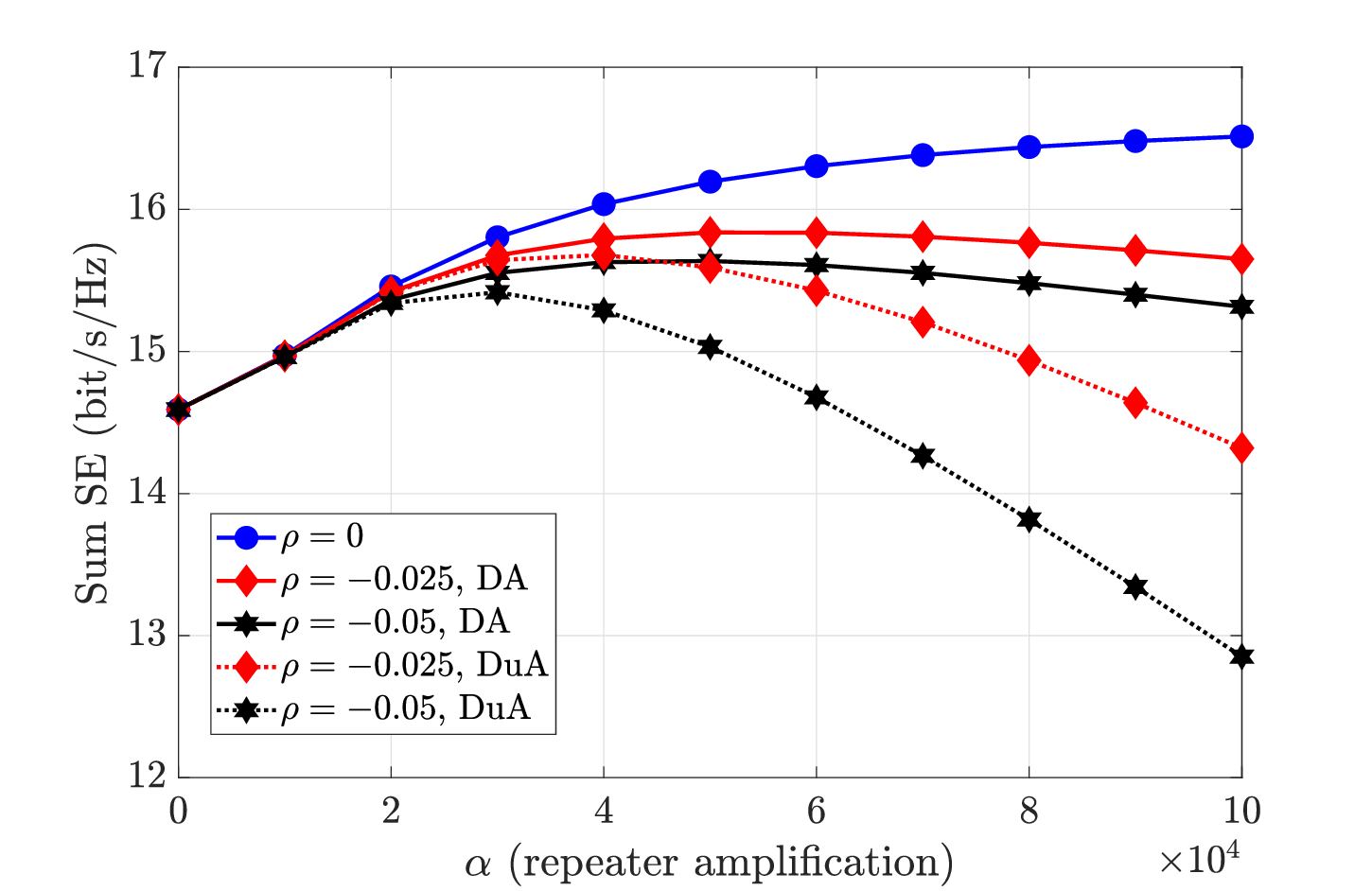}
\end{overpic} 
\vspace{-5mm}
        \caption{Sum SE in terms of repeater amplification gain $\alpha$ for $K=8$, when the repeater-BS distance is $200$\,m and UEs occupy a $100\,\text{m}\times100\,\text{m}$ region. }
        \label{fig2}
        \vspace{-4mm}
\end{figure}

In Fig.~\ref{fig2}, we increase the number of UEs to $K=8$ and repeat the experiment from the previous figure. As $K$ grows, multiuser interference becomes stronger and the non-linear distortion accumulates, which makes DA combining increasingly critical. Beyond a certain amplification level, the performance of the DuA combiner degrades sharply and even falls below the no-repeater baseline. The DA combiner also exhibits an optimal amplification point, but it remains substantially more robust to increasing $\alpha$ compared to DuA.

\begin{figure}[t!]
        \centering
	\begin{overpic}[width=0.98\columnwidth,trim=0.2cm 0cm 0.5cm 0.5cm,clip,tics=10]{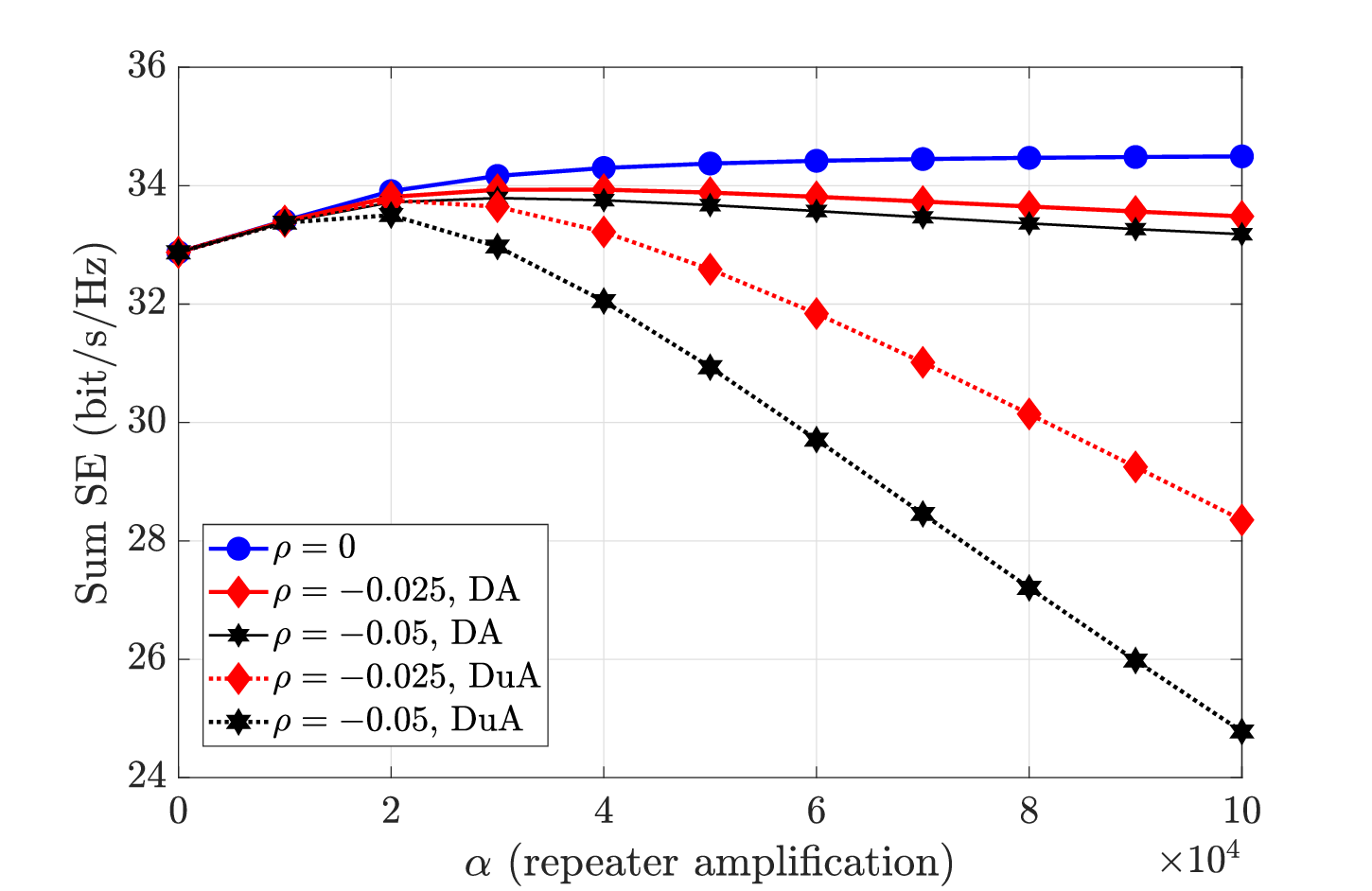}
\end{overpic} 
\vspace{-5mm}
        \caption{Sum SE in terms of repeater amplification gain $\alpha$ for $K=8$, when the repeater-BS distance is $100$\,m and UEs occupy a $100\,\text{m}\times100\,\text{m}$ region. }
        \label{fig3}
        \vspace{-4mm}
\end{figure}

In Fig.~\ref{fig3}, we move the repeater closer to the BS—placing it 100\,m to the right of the BS (i.e., at \((100,0)\))—and uniformly deploy the UEs within
\vspace{-2mm}
\begin{align}
(x,y)\in [100,200]\times[-50,50]\ \text{meters}.
\end{align}
The number of UEs is $K=8$. The stronger repeater–BS link (higher channel gain) causes the repeater to forward non-linear distortion more aggressively. Consequently, DuA combining exhibits a marked SE loss. While DA combining remains comparatively robust, the incremental benefit of using a repeater is limited in this geometry: because the UEs are already close to the BS, the direct links are strong and the no-repeater baseline attains high SE, leaving little headroom for repeater-assisted gains.

\vspace{-3mm}

\section{Conclusions}

\vspace{-2mm}

We studied uplink RA-MIMO with a single-antenna NCR subject to PA nonlinearities and developed a DA receive design grounded in a Bussgang linear–distortion model. Closed-form expressions for the Bussgang gain and distortion covariance enabled an achievable SE characterization and the design of a DA combiner that maximizes the signal-to-interference-plus-distortion ratio.

Our simulations across amplification gains $\alpha$, non-linearity levels $\rho$, UE number $K$, and deployment geometries demonstrate three key insights. First, DA processing is essential: by explicitly modeling distortion statistics, it preserves repeater-assisted SE gains over broad $(\alpha,\rho)$ regimes, whereas DuA combining can degrade sharply with increasing $\alpha$ or $K$, even underperforming the no-repeater baseline. Second, the results reveal a strong operating-point sensitivity: while moderate amplification improves SE, excessive gains forward distortion more aggressively, making the choice of $\alpha$ critical. Third, the deployment geometry plays a decisive role: when the repeater is close to the BS, the forwarded distortion dominates and limits the benefit of repeater-assisted operation; in more moderate geometries, DA-enabled repeater operation yields clear and robust SE improvements.

\bibliographystyle{IEEEbib}
\bibliography{strings,refs}

\end{document}